\documentclass{article}
\usepackage{ijcai25}

\usepackage{times}
\usepackage{soul}
\usepackage{url}
\usepackage[hidelinks]{hyperref}
\usepackage[utf8]{inputenc}
\usepackage[small]{caption}
\usepackage{graphicx}
\usepackage{amsmath}
\usepackage{amsthm}
\usepackage{booktabs}
\usepackage{algorithm}
\usepackage{algorithmic}
\usepackage[switch]{lineno}

\usepackage{bm}
\usepackage{xcolor}
\usepackage{amsthm}
\usepackage{amsmath}
\usepackage{amssymb}
\usepackage{booktabs}
\usepackage{graphicx}
\usepackage{enumitem}
\usepackage{mathtools}
\usepackage{microtype}
\usepackage{subfigure}
\usepackage[normalem]{}
\usepackage[page]{appendix}
\usepackage{threeparttable}
\usepackage{multirow, makecell}
\usepackage[textsize=tiny]{todonotes}

\usepackage{comment}

\title{InstGAN: Instant Actor-Critic-Driven GAN for De Novo Molecule Generation and Property Optimization}

\author{
Huidong Tang$^{1,2}$
\and
Chen Li$^{3*}$\and
Sayaka Kamei$^2$\and
Yoshihiro Yamanishi$^{4}$\And
Yasuhiko Morimoto$^2$\\
\affiliations
$^1$Cloud Computing Technology and Application Department, Shandong Institute of Commerce and Technology, Jinan, China \\
$^2$Graduate School of Advanced Science and Engineering, Hiroshima University, Higashi-Hiroshima, Japan \\
$^3$D3 Center, The University of Osaka, Osaka, Japan \\
$^4$Graduate School of Informatics, Nagoya University, Nagoya, Japan \\
\emails
tanghd24@163.com,
li.chen.d3c@osaka-u.ac.jp,
s10kamei@hiroshima-u.ac.jp,
yamanishi@i.nagoya-u.ac.jp,
morimo@hiroshima-u.ac.jp
}

\begin{document}

\maketitle

\begin{abstract}
    Deep generative models, such as generative adversarial networks (GANs), have been employed for $de~novo$ molecular generation in drug discovery. Most prior studies have utilized reinforcement learning (RL) algorithms, particularly Monte Carlo tree search (MCTS), to handle the discrete nature of molecular representations in GANs. However, due to the inherent instability in training GANs and RL models, along with the high computational cost associated with MCTS sampling, MCTS RL-based GANs struggle to scale to large chemical databases. To tackle these challenges, this study introduces a novel GAN based on actor-critic RL with instant and global rewards, called InstGAN, to generate molecules at the token-level with multi-property optimization. Furthermore, maximized information entropy is leveraged to alleviate the mode collapse. The experimental results demonstrate that InstGAN outperforms other baselines, achieves comparable performance to state-of-the-art models, and efficiently generates molecules with multi-property optimization. The code is available at: \url{https://github.com/tang777777/InstGAN}.
\end{abstract}

\section{Introduction}
\label{sec:intro}
Modern human healthcare and well-being are intricately intertwined with the field of drug discovery, which seeks to uncover new chemical compounds with therapeutic effects. However, traditional drug discovery is a time-consuming and expensive endeavor, taking an average of 12 years and costing 2.6 billion USD \cite{chan2019advancing}. To expedite the process and mitigate costs, artificial intelligence (AI) has garnered the attention of the pharmaceutical industry \cite{paul2021artificial}. Among the recent applications of AI, deep generative models have demonstrated remarkable progress, as exemplified by DALL$\cdot$E2 in the realm of computer vision and ChatGPT in natural language processing (NLP) \cite{openai2023gpt}. The adoption of such models has also become increasingly prominent in the field of drug discovery \cite{chen2018rise}.

Molecular graphs \cite{shi2020graphaf} and simplified molecular input line entry systems (SMILES) strings \cite{weininger1988smiles} constitute the two primary representations of molecules in deep generative models. However, generating molecules with desired chemical properties using such discrete representations is a non-trivial task. Most prior studies related to generative adversarial networks (GANs) \cite{yu2017seqgan,guimaraes2017objective,de2018molgan} typically update the generator by integrating the output probability of the discriminator with the chemical properties of generated molecules as a reward for reinforcement learning (RL), following the REINFORCE algorithm \cite{williams1992simple}. Due to the inability of GANs to calculate rewards for partially generated molecules, Monte Carlo tree search (MCTS) is frequently utilized for sampling and completing molecules \cite{li22transformer}. Unfortunately, the integration of RL algorithms with GANs further exacerbates the instability of the training process. Achieving training stability with MCTS demands a substantial number of samples, rendering the process highly time-consuming \cite{li2023spotgan}.

Furthermore, most aforementioned studies on $de~novo$ molecular generation have focused on optimizing a single chemical property. However, in practical applications, it is often desirable to generate molecules that satisfy multiple chemical property constraints. In contrast to the former, multi-property optimization is highly complicated and challenging to achieve in nature. This is because the multi-property optimization task entails not only learning the semantic and syntactic rules of molecules to generate valid molecules from scratch but also finding pathways to optimize the distribution of chemical properties during the process \cite{barshatski2021multi}. For example, molecules exhibiting both drug-likeness and dopamine receptor (DRD2) activity represent only $1.6\%$ of the generated molecules \cite{jin2019hierarchical}. Therefore, employing deep generative models for $de~novo$ molecular generation with multi-property optimization holds significance to the drug discovery industry \cite{barshatski2021unpaired}.

Inspired by the previous studies in \cite{de2018molgan,tang2023earlgan}, this study introduces a novel GAN based on actor-critic RL with instant rewards (IR) and global rewards (GR), called InstGAN, to generate molecules at the token-level with multi-property optimization. Specifically, the generator is constructed using a long short-term memory (LSTM) 
that generates SMILES strings in an autoregressive manner. The discriminator quantifies each token based on SMILES substrings produced by the generator. To enhance the ability of the discriminator to quantize tokens, a bidirectional LSTM (Bi-LSTM) is chosen as the discriminator. Additionally, multi-property prediction networks with the same structure as the discriminator predict the corresponding property scores for each token as well. Subsequently, the scores of discriminator and property prediction networks, along with their global-level scores, serve as rewards for RL. To expedite the training process and facilitate its application to extensive chemical databases, an IR calculation based on actor-critic RL is proposed to update the generator. Furthermore, the maximized information entropy (MIE) is included in the generator loss function to mitigate mode collapse and enhance molecular diversity. The main contributions are 
\begin{itemize}
    \item{\bf Novel reward calculation:} This study proposes an actor-critic RL-based approach to calculate IR and GR for molecular generation with multi-property optimization.
    \item{\bf Scalability for chemical property optimization:} InstGAN exhibits versatile scalability, enabling flexible expansion from single-property to arbitrary multi-property optimized molecular GAN.
    \item{\bf Superior performance:} Experimental results validate that InstGAN outperforms other baselines, achieves comparable performance to state-of-the-art (SOTA) models, and demonstrates the ability to generate molecules with multi-property optimization in a fast and efficient manner.
\end{itemize}

\section{Related Work}
\label{sec:related}
\subsection{Variational Autoencoder (VAE)-based Models}
\label{subsec:vae_models}
Two VAE variants, Character-VAE and Grammar-VAE \cite{kusner2017grammar}, integrate parse trees with VAEs to facilitate the generation of syntactically valid molecules. However, due to ignoring the semantic information of the molecular representations, the correlation between the training set and the generated molecules cannot be guaranteed. In contrast, Syntax-VAE \cite{dai2018syntax} ensures that the generated molecules are both syntactically valid and semantically meaningful. JT-VAE \cite{jin2018junction} adopts a two-step VAE approach, first generating tree-structured scaffolds based on chemical substructures and then combining these outputs into complete molecules using a graph message-passing network \cite{dai2016discriminative,gilmer2017neural}. Nonetheless, a major limitation of VAEs is the typically limited size of the latent space, which may restrict the capacity to produce highly realistic molecules.

\subsection{Flow-based Models}
\label{subsec:flow_models}
GraphAF \cite{shi2020graphaf} utilizes a flow-based autoregressive model to generate molecular graphs. GraphDF \cite{luo2021graphdf} incorporates invertible modulo shift transforms to connect discrete latent variables with graph nodes and edges, resulting in the generation of molecular graphs. MoFlow \cite{zang2020moflow} adopts a Glow-based model \cite{kingma2018glow} to generate chemical bonds as graph edges and employs a graph conditional flow to subsequently generate atoms as graph nodes, followed by posthoc validity correction. GraphCNF \cite{lippe2021categorical} falls under the category of flow-based models, commonly applied in various data domains. In molecular graph generation, GraphCNF leverages flow-based techniques to address the unique challenges associated with generating molecular graphs. However, flow-based models 
exhibit several notable limitations: the computation of the Jacobian matrix is time-consuming, often requiring approximations, and the necessity for network invertibility imposes constraints on their representational capacity, limiting their flexibility in modeling complex, high-dimensional data distributions \cite{zhang2021iflow}. Additionally, ensuring invertibility across all layers can lead to challenges in model training and optimization.

\subsection{Diffusion-based Models}
\label{subsec:diffusion_models}
Recently, diffusion models have found application in the domain of molecular generation, where they are being utilized to tackle the intricate challenges associated with generating molecular structures that adhere to specific chemical and property constraints. DiGress \cite{vignac2023digress} iteratively refines noisy graphs by adding or removing edges and adjusting categories, which results in the generation of molecular graphs with node and edge attributes suitable for classification. GDSS \cite{jo2022score} and D2L-OMP \cite{guo2023diffusing} are the two SOTA models in the realm of molecular generation. GDSS skillfully captures the joint distribution of molecular nodes and edges, generating molecular replicas closely aligned with the training distribution. D2L-OMP generates molecules with property optimization based on a hybrid Gaussian distribution by employing diffusion on two structural levels: molecules and molecular fragments. This hybrid Gaussian distribution is then utilized in the reverse denoising process.

\subsection{GAN-based Models}
\label{subsec:gan_models}
SeqGAN \cite{yu2017seqgan} pioneered the incorporation of MCTS-based RL into GAN architecture, specifically designed to handle discrete text. This innovation has inspired various studies on molecular generation using GANs. MolGAN \cite{de2018molgan} introduces an implicit, likelihood-free discrete GAN for generating small molecular graphs. However, MolGAN faces an overfitting problem, leading to less than 5\% uniqueness in the generated molecules. ORGAN \cite{guimaraes2017objective} integrates domain-specific knowledge as rewards for generating SMILES strings through the MCTS-based RL algorithm. TransORGAN \cite{li22transformer} leverages a transformer architecture to capture semantic information and employs variant SMILES technique for syntax rule learning. SpotGAN \cite{li2023spotgan} adopts a first-decoder-then-encoder transformer model for generating SMILES strings from a given scaffold. However, the use of MCTS-based RL in GANs often demands a substantial number of samples for training stability, making it impractical for extensive chemical databases. EarlGAN \cite{tang2023earlgan}, while capable of generating valid molecules on large chemical databases, lacks the ability to optimize chemical properties, particularly multiple properties simultaneously. 

This study introduces an actor-critic RL-driven GAN that employs IR and GR to enhance the efficiency of learning semantic and syntactic rules in SMILES strings. Furthermore, our other goal is to optimize molecule generation across diverse chemical properties.

\section{Model}
\label{sec:model}
\begin{figure*}[t]
\centering
\includegraphics[width=1\textwidth]{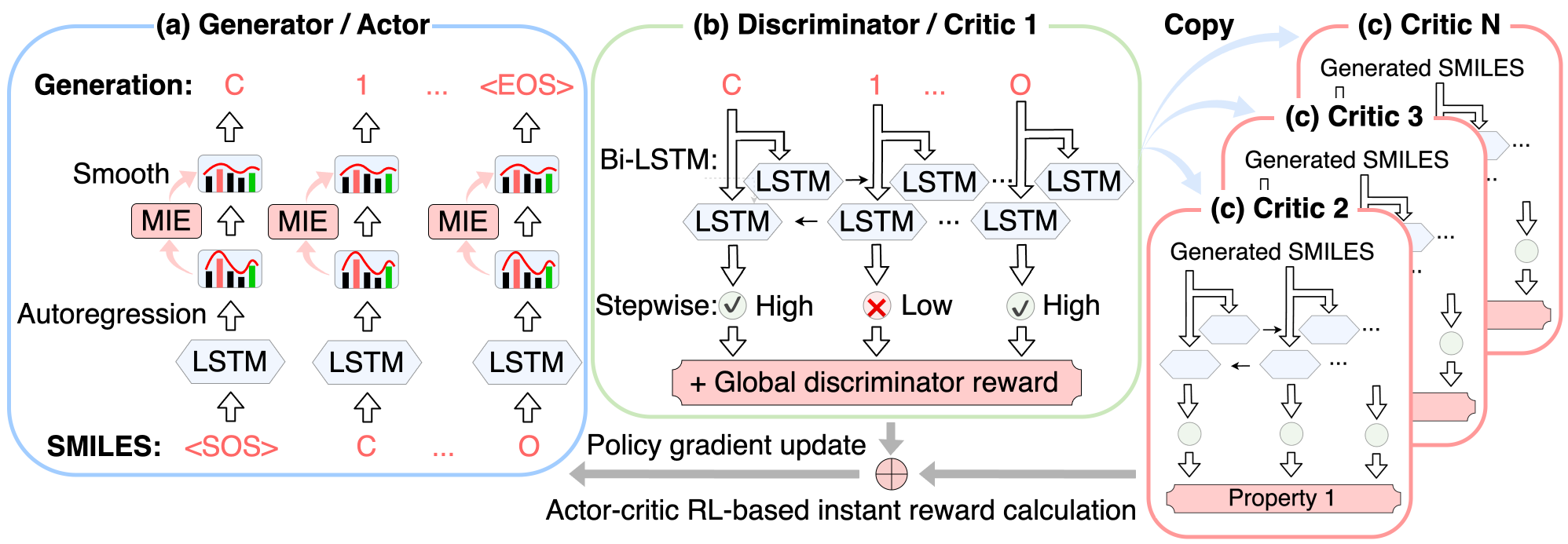}
\caption{The architectural overview of InstGAN comprises three main substructures. \textbf{(a) The generator}, featuring an LSTM, produces tokens in an autoregressive manner at each time step. MIE is employed to enhance the likelihood of sampling different tokens, smoothing the output probability distribution for the generator. \textbf{(b) The discriminator}, a Bi-LSTM, scores generated tokens based on both forward and backward at each time step, enabling semantic and syntactic discrimination at the token-level. Higher probabilities are assigned to likely generated tokens, while those with errors receive lower probabilities. The average of all-token probabilities assesses overall generation quality from a global perspective. \textbf{(c) Multiple pretrained chemical property prediction networks}, labeled Critic $1$ to Critic $N$, have the same structure as the discriminator. They predict various chemical properties for each token of a generated SMILES string. Similarly, the sum of stepwise scores is averaged to provide a global property score for the entire SMILES string. Finally, scores from the discriminator and critics serve as rewards in the actor-critic RL algorithm, co-updating the generator via the policy gradients.}
\label{fig:model}
\end{figure*}

Figure \ref{fig:model} illustrates the overall architecture of InstGAN and highlights the three key substructures (i.e., the generator, discriminator, and multiple chemical property prediction networks) that are crucial for generating molecules with multi-property optimization from scratch. Formally, let $G_{\bm\theta}$ and $D_{\bm\phi}$ represent the generator and discriminator of InstGAN with parameters $\bm\theta$ and $\bm\phi$, respectively. ${\bm S}_{1:T}=[s_1,\cdots,s_T]$ denotes a SMILES string with a sequence length of $T$. Then, the min-max optimization procedure \cite{goodfellow2014generative} is implemented as follows:
\begin{align}
&\min_{\bm\theta}\max_{\bm\phi} V(G_{\bm\theta}, D_{\bm\phi})= \nonumber\\
&\mathbb{E}_{{\bm S} \sim p_r({\bm S})}[\log D_{\bm\phi} ({\bm S})] + \mathbb{E}_{{\bm S} \sim p_{\bm\theta}({\bm S})}[\log (1 - D_{\bm\phi}({\bm S}))],
\end{align}
where $p_r(\cdot)$ and $p_{\bm\theta}(\cdot)$ represent the distributions of training SMILES strings and generated sequences, respectively. Following a similar approach to NLP methods \cite{de2019training,fedus2018maskgan}, InstGAN utilizes an autoregressive generator and discriminator. Notably, this design allows for the allocation of dense rewards at the token-level. For detailed explanation of InstGAN's design motivations and a full description of its generator and discriminator architectures, see Appendix A\footnote{The Appendix is available at: https://github.com/tang777777/InstGAN/blob/main/Appendix.pdf}.

\subsection{Token-level Critics}
\label{subsec:token}
Unlike graph-based approaches, molecular generative models relying on SMILES strings often face challenges in ensuring high validity due to the intricacies of checking valence during the autoregressive generation process. Typically, invalid SMILES strings may arise from mismatched tokens, requiring the removal or replacement of other suitable tokens to restore validity. A crucial aspect of addressing this issue lies in the meticulous assessment of SMILES strings at the token level. We therefore integrate the validity assessment directly into the generation process, by treating each token generation as an action in a token-level actor-critic RL loop: at every step, the generator (actor) proposes a candidate token, while the discriminator (critic) assess its syntactic validity and return a dense reward. This feedback system ensures the generator of high-validity token generation.

\paragraph{Token-level discriminator.} In InstGAN, the generator employs an autoregressive process to produce SMILES strings, serving as inputs for the discriminator. Diverging from traditional discriminators, InstGAN performs token-level discrimination for each generated token. Specifically, with a Bi-LSTM, let $\overrightarrow{\bm S}_{1:t}$ and $\overleftarrow{\bm S}_{t:T}$ represent the forward and backward SMILES substrings, respectively. The discriminator calculates the probability $r^D_t$ that it deems the $t$-th token $\widetilde{s}_t$ of the SMILES string as true. The calculation is as follows:
\begin{align}
&r^D_t = D_{\bm\phi} (\widetilde{s}_t | \overrightarrow{\bm S}_{1:t}, \overleftarrow{\bm S}_{t:T}). 
\end{align}
To assess the validity of the entire SMILES string, we calculate the global discriminator score as
\begin{align}
&r^D_{1:T} = \frac{1}{T}\sum_{t=1}^T r^D_t.
\end{align}

\paragraph{Token-level property prediction networks.} The chemical property prediction networks are replicated from the discriminator and share the same structure. In the pre-training phase, the entire SMILES string serves as input, and the networks utilize the corresponding property values of the complete SMILES string as labels for the chemical properties of its tokens at each step. In the training phase of InstGAN, these networks calculate both instant and global chemical properties of the generated SMILES string. Collaborating with the discriminator, they contribute to updating the generator using actor-critic RL-based rewards. 

\subsection{Instant and Global Reward Calculation}
\label{subsec:reward}
Following the actor-critic RL algorithm, the generator functions as the actor network responsible for action selection, while the discriminator, along with the property prediction networks, serves as the critic for reward calculation. However, in contrast to the traditional actor-critic RL algorithm \cite{konda1999actor}, which calculates the reward at the last time step $T$, we compute the reward for each token as
\begin{align}
&R^D_t = 2 r^D_t - 1 + W^D r^D_{1:T},
\end{align}
where $W^D$ represents the weight assigned to the discriminator's GR. Similarly, the reward of the property prediction networks can be calculated and denoted as $R^{C_i}_t$, $i \in [1, 2, \cdots, N]$, and $N$ indicates the number of chemical properties to be optimized. Subsequently, the overall reward for chemical properties is represented as $R^C_t = \sum_{i=1}^N W_{C_i} R^{C_i}_t$, where
$W_{C_i}$ denotes the weight assigned to the $i$-th chemical property prediction network. Finally, the total reward $R_t$ 
can be calculated as follows:
\begin{align}
&R_t = (1-\lambda) R^D_t + \lambda R^C_t,
\end{align}
where $\lambda$ represents a hyperparameter that balances the trade-off between the GAN and RL.

\begin{algorithm}[t]
\caption{Training Procedure of InstGAN.}
\label{alg:algorithm1}
\begin{algorithmic}[1]

\STATE {\bfseries Data:} a SMILES dataset ${\mathcal{D}}_{real}$
\STATE {\bfseries Initialization:} $G_{\bm\theta}$, $D_{\bm\phi}$, $C^i_{{\bm\varphi}_i}, i \in [1, \cdots, N]$ 

\STATE Generate a dataset ${\mathcal{D}}_{fake}$  from scratch

\STATE // {\ttfamily Pre-train the discriminator}
\FOR{$k=1 \to \text{d-steps}$}
\STATE Update $D_{\bm\phi}$ on ${\mathcal{D}}_{real}$ and ${\mathcal{D}}_{fake}$
\ENDFOR

\STATE // {\ttfamily Pre-train the generator}
\FOR{$k=1 \to \text{g-steps}$}
\STATE Update $G_{\bm\theta}$ on ${\mathcal{D}}_{real}$
\ENDFOR

\STATE // {\ttfamily Pre-train property networks}
\FOR{$i=1 \to N$}
    \FOR{$k=1 \to \text{p-steps}$}
    \STATE Update $C^i_{{\bm\varphi}_i}$ on ${\mathcal{D}}_{real}$
    \ENDFOR
\ENDFOR

\STATE // {\ttfamily Multi-property optimization}
\FOR{$k=1 \to \text{m-steps}$}
\STATE $G_{\bm\theta}$ updates the generated dataset ${\mathcal{D}}_{fake}$
\STATE Update $D_{\bm\phi}$ and $C^i_{{\bm\varphi}_i}$
\STATE $D_{\bm\phi}$ discriminates between ${\mathcal{D}}_{real}$ and ${\mathcal{D}}_{fake}$ and outputs the IR and GR
\STATE $C^i_{{\bm\varphi}_i}$ calculates the IR and GR for chemical properties
\STATE Update $G_{\bm\theta}$ based on the rewards
\ENDFOR
\end{algorithmic}
\end{algorithm}

\subsection{Objective Function}
\label{subsec:objective}
In the iterative adversarial process, the generator is updated by the MCTS-based RL algorithm through the sampling of numerous samples, leading to a computational training process. In contrast, InstGAN utilizes rewards derived from the actor-critic RL algorithm and MIE for the calculation. Please see the description of the MCTS-based RL and actor-critic RL for algorithms comparison in Appendix B. The overall loss function for the generator is calculated as
\begin{align}
&\mathcal{L}_{\bm \theta} = \mathcal{L}_{RL} + \beta \mathcal{L}_{MIE},
\end{align}
where $\mathcal{L}_{RL}$ and $\mathcal{L}_{MIE}$ represent the loss functions of RL and MIE, respectively, with $\beta$ serving as the trade-off parameter between them. In accordance with the policy gradient, $\mathcal{L}_{RL}$ is calculated by minimizing the expected reward score:
\begin{align}
&\mathcal{L}_{RL} = - \frac{1}{T} \sum_{{\bm S}_{1:T}} (R_t - b_t) \log p_{\bm\theta}(s_t | {\bm S}_{1:t-1}),
\end{align}
where $b_t$ denotes the baseline using the global moving-average rewards \cite{sutton2018reinforcement}, which is calculated using both the mean reward $\Bar{R}$ across the current batch and the previous baseline $b_{t-1}$, calculated as
\begin{align}
&b_t = (1 - \alpha) \Bar{R} + \alpha b_{t-1}.
\end{align}
Here, $\alpha$ denotes a smoothing factor. To encourage the generator to sample tokens with probabilities other than the highest, MIE $\mathcal{L}_{MIE}$ is incorporated into the generator's loss function. This addition serves to smooth the probability distribution, mitigating the mode collapse problem and promoting diversity in generating molecules.
\begin{align}
&\mathcal{L}_{MIE} = \frac{1}{T} \sum_{{\bm S}_{1:T}} \sum_{v=1}^{V} p_{\bm\theta}(s^v_t) \log p_{\bm\theta}(s^v_t),
\end{align}
where $V$ represents the size of chemical vocabulary. Algorithm \ref{alg:algorithm1} outlines the training procedure for InstGAN. The generator, discriminator, and multiple chemical property prediction networks are first pre-trained. Then, the generator is trained to generate a dataset. Afterward, the discriminator and critics are updated using both real and generated SMILES strings. Finally, the IR and GR values are calculated to update the generator's parameters.

\section{Experiments}
\label{sec:exp}
\subsection{Experimental Setup}
\label{subsec:setup}
\paragraph{Datasets.} The performance of InstGAN was validated through experiments on two widely used chemical datasets: ZINC \cite{irwin2012zinc} and ChEMBL \cite{gaulton2017chembl}. The ZINC dataset contains $250,000$ drug-like molecules. The ChEMBL dataset includes approximately 1.6 million molecules, with each having a median of 27 and a maximum of 88 heavy atoms.

\paragraph{Chemical properties.}
\textbf{Drug-likeness (QED)} quantifies the probability that a molecule belongs to a drug \cite{bickerton2012quantifying}.
\textbf{Solubility (logP)} measures how well a molecule dissolves in lipid versus aqueous environments, quantified as the Log of the partition coefficient \cite{comer2001lipophilicity}.
\textbf{Synthesizability (SA)} is defined by the synthetic accessibility score, evaluating the ease with which a molecule can be synthesized \cite{ertl2009estimation}. 
\textbf{Dopamine receptor D2 (DRD2)} is a central nervous system G protein-coupled receptor that is essential for dopamine-mediated signaling \cite{olivecrona2017molecular}. Table \ref{table:statistic} provides detailed statistical descriptions of the datasets.

\paragraph{Metrics.} 
\begin{table}
\centering
\setlength\tabcolsep{0.6pt}
\begin{threeparttable}
\begin{tabular}{cccccccc}
\hline
Dataset & MaxL & MinL & AvgL & QED & logP & SA & DRD2 \\\hline
ZINC&109 & 9 & 44 & 0.73 & 0.56 & 0.56 & 0.24 \\
ChEMBL& 116 & 10 & 47 & 0.57 & 0.67 & 0.62 & 0.25 \\\hline
\end{tabular}
\begin{tablenotes}
\footnotesize
\item[$\star$] MaxL, MinL, and AvgL indicate the maximum, minimum, and average length of the SMILES strings.
\end{tablenotes}
\end{threeparttable}
\caption{Statistical descriptions of the datasets.}
\label{table:statistic}
\end{table}

\textbf{Validity} is assessed by the proportion of chemically valid molecules among all generated ones, validated practically using the RDKit tool \cite{landrum2013rdkit}. \textbf{Uniqueness} is determined by the proportion of non-duplicated molecules among all valid ones. \textbf{Novelty} is defined as the proportion of unique molecules not present in the training set. \textbf{Total} is the ratio of novel molecules to all generated ones. \textbf{Diversity} is calculated as the average Tanimoto distance \cite{rogers1960computer} between the Morgan fingerprints \cite{cereto2015molecular} of novel molecules. All these properties and metrics are normalized to a range of $[0, 1]$, with a higher score indicating better performance. For a detailed description of the chemical properties, evaluation metrics and hyperparameters, please refer to Appendix C.

\begin{table*}
\centering
\resizebox{0.82\textwidth}{!}{
\begin{threeparttable}
\centering
\begin{tabular}{clcccc}\toprule
& Model & Validity (\%) $\uparrow$ & Uniqueness (\%) $\uparrow$ & Novelty (\%) $\uparrow$ & Total (\%) $\uparrow$ \\\hline
\multirow{4}{7em}[-8pt]{\makecell{VAE-based}}
& RNN-Attention \cite{dollar2021attention} & 71.57 & 99.94 & \colorbox[gray]{0.9}{100.0} & 71.53 \\
& TransVAE \cite{dollar2021attention} & 25.39 & 99.96 & \colorbox[gray]{0.9}{100.0} & 25.38 \\
& Character-VAE \cite{kusner2017grammar} & 86.65 & 81.21 & 26.36 & 18.55 \\
& Grammar-VAE \cite{kusner2017grammar} & 91.91 & 77.24 & 11.90 & 8.45 \\
& JT-VAE \cite{jin2018junction} & \colorbox[gray]{0.9}{100.0} & 19.75 & 99.75 & 19.70 \\\hline

\multirow{4}{7em}[-2pt]{\makecell{Flow-based}}
& GraphAF \cite{shi2020graphaf} & 68.00 & 99.10 & \colorbox[gray]{0.9}{100.0} & 67.39 \\
& GraphDF \cite{luo2021graphdf} & 89.03 & 99.16 & \colorbox[gray]{0.9}{100.0} & 88.28 \\
& MoFlow \cite{zang2020moflow} & 81.76 & 99.99 & \colorbox[gray]{0.9}{100.0} & 81.75 \\
& GraphCNF \cite{lippe2021categorical} & 63.56 & \colorbox[gray]{0.9}{100.0} & \colorbox[gray]{0.9}{100.0} & 63.56 \\\hline

\multirow{4}{7em}[11pt]{\makecell{Diffusion-based}}
& GDSS \cite{jo2022score} & 97.01 & 99.64 & \colorbox[gray]{0.9}{100.0} & 96.66 \\
& D2L-OMP \cite{guo2023diffusing} & 97.51 & 99.88 & \colorbox[gray]{0.9}{100.0} & \colorbox[gray]{0.9}{97.39} \\\hline

\multirow{4}{7em}[-2pt]{\makecell{GAN-based}}
& ORGAN \cite{guimaraes2017objective} & 67.96 & 98.20 & 98.39 & 65.66 \\
& MolGAN \cite{de2018molgan} & 95.30 & 4.30 & \colorbox[gray]{0.9}{100.0} & 4.10 \\
& TransORGAN \cite{li22transformer} & 74.31 & 91.79 & \colorbox[gray]{0.9}{100.0} & 68.21 \\
& SpotGAN \cite{li2023spotgan} & 93.26 & 92.78 & 92.75 & 80.25 \\\hline

\multirow{4}{7em}[-7pt]{\makecell{\textbf{InstGAN}}}
& Pre-train (Average) & 95.45 & 98.63 & 99.71 & 93.87 \\
& Property (QED) & 97.89 & 98.31
 & 99.69
 & 95.94 \\
& Property (logP) & 96.65
 & 98.42
 & 99.93 & 95.05 \\
& Property (SA) & 97.46
 & 98.59
 & 99.75
 & 95.85 \\
& Multi-property & 97.71 & 98.71 & 99.64 & 96.10 \\ \toprule
\end{tabular}
\begin{tablenotes}
\item[$\star$] The values in the gray cells indicate the maximum scores in the respective columns.
\end{tablenotes}
\end{threeparttable}
}
\caption{Comparison results of InstGAN with various baseline models for chemical property optimization on the ZINC dataset.}
\label{table:compare_res}
\end{table*}

\subsection{Evaluation Results}
\label{subsec:compare}
\paragraph{Comparison results with baselines.} Table \ref{table:compare_res} compared the results of InstGAN with various baseline models (including VAE-, flow-, diffusion-, and GAN-based models) for chemical property optimization on the ZINC dataset. To ensure a fair comparison, InstGAN was pre-trained several times, and the average results are presented. Additional details on these multiple pre-training sessions are provided in Appendix D. For VAE-based models, although RNN-Attention and TransVAE generated molecules that were highly unique and novel, their validity (i.e., $<71.6\%$) was significantly lower compared to InstGAN. InstGAN outperformed Character-VAE and Grammar-VAE in all metrics. Although JT-VAE exhibited high validity and novelty, the uniqueness was only $19.75\%$, significantly lower than that of InstGAN. For flow-based models, despite exhibiting high uniqueness and novelty, their validity was lower, specifically less than $89.03\%$, which significantly trails behind InstGAN. InstGAN demonstrated comparable performance to SOTA diffusion models GDSS and D2L-OMP. All models, including pre-training and tasks involving single properties (QED, logP, SA), as well as multi-property tasks (QED, logP, SA), achieved an overall score surpassing 93.87\%. Among GAN-based models, while MolGAN and TransORGAN exhibited a novelty of 100.0\%, their validity and uniqueness fell significantly lower than InstGAN. Specifically, MolGAN demonstrated a mere 4.3\% uniqueness, and TransORGAN exhibited a validity of only 74.31\%. ORGAN's validity stood at 67.96\%, markedly lower than InstGAN's impressively high validity exceeding 95.45\%. Furthermore, InstGAN outperformed SpotGAN in terms of validity, uniqueness, novelty, and total score, with the highest overall performance. InstGAN was trained for single- and multi-property optimization. In single-property optimization, InstGAN was trained individually using QED, logP, and SA. In multi-property optimization, these three properties were used to jointly train InstGAN. The validity, uniqueness, novelty, and total score all reached up to $93.87\%$. Overall, while InstGAN generated molecules from less informative SMILES strings compared to graph-based models, it surpassed VAE-, flow-, and GAN-based baselines and demonstrated comparable performance to SOTA diffusion models. This highlights InstGAN's robust capability in molecular generation, demonstrating excellence in both single-property and multi-property optimization.

\paragraph{Property optimization.} Owing to the diverse sequence representations in SMILES strings, they inherently introduce more noise compared to molecular graphs. Furthermore, as molecular graphs typically contain more detailed information, including atoms, chemical bonds, and valences, the task of molecular generation based on SMILES strings is more challenging. As illustrated in Table \ref{table:compare_res}, InstGAN performed well in molecular generation in all evaluation metrics. InstGAN obtained almost $100\%$ novelty in multi-property optimization, with validity and uniqueness exceeding $97.7\%$. InstGAN's ability to excel in learning both semantic and syntactic features within SMILES strings is the key contributing factor to this achievement. This achievement surpassed the capabilities of all other models that rely on SMILES strings for chemical property optimization.

\begin{table}
\setlength\tabcolsep{0.35pt}
\centering
\begin{tabular}{lcccccc}\toprule
Model (Property) & Top-1 & Top-5 & Top-10 & Top-100 & Top-1000 \\\hline
GraphAF & 0.94 & 0.93 & 0.92 & 0.86 & 0.57 \\
GDSS & 0.94 & 0.94 & 0.93 & 0.91 &  0.85 \\
MoFlow & 0.93 & 0.92 & 0.92 & 0.87 & 0.78 \\
D2L-OMP & \colorbox[gray]{0.9}{0.95} & 0.94 & 0.94 & 0.91 & 0.85 \\\hline
\textbf{InstGAN} & \colorbox[gray]{0.9}{0.95} & \colorbox[gray]{0.9}{0.95} & \colorbox[gray]{0.9}{0.95} & \colorbox[gray]{0.9}{0.95} & \colorbox[gray]{0.9}{0.94} \\
\textbf{InstGAN} (Multi) & \colorbox[gray]{0.9}{0.95} & \colorbox[gray]{0.9}{0.95} & \colorbox[gray]{0.9}{0.95} & 0.94 & 0.93 \\
\hline
\end{tabular}
\caption{QED assessment of the top-k generated molecules.}
\label{table:top_k_res}
\end{table}

Table \ref{table:top_k_res} and Table E.1 show the top-k property scores for the multi- and single-property optimization of the generated molecules, respectively. In multi-property optimization, InstGAN enhanced all targeted chemical properties. The QED scores showed a 30.1\% improvement (from 0.73 to 0.95) for Top-1 and 27.4\% improvement (from 0.73 to 0.93) for Top-1000, compared with the training dataset. Furthermore, the generated molecules of InstGAN with logP and SA as the desired properties exhibited a notable improvement, with logP scores increasing by 78.6\% for Top-1 and 58.9\%, as well as 76.8\% for SA scores for Top-1000, respectively. In both property optimization, InstGAN generated molecules with higher QED scores, compared with other baselines. Especially, InstGAN improved the QED score of the Top-1000 by 9.4\% and 10.6\%, comparing to the SOTA D2L-OMP baseline. In single-property optimization, the generated molecules of InstGAN demonstrated substantial improvements, with the scores increasing by 30.1\%, 78.6\%, and 78.6\% for Top-1, and 28.8\%, 78.6\%, and 78.6\% for Top-1000, compared with the training dataset. Overall, InstGAN showcased substantial enhancements in property optimization, underscoring its effectiveness in improving targeted chemical properties.

Figures E.1 and E.2 in the appendix show the top-ranked (Top-1) molecular structures generated by InstGAN in the single- and multi-property optimization tasks, respectively. The generated molecule adhered to H{\"u}ckel's rules \cite{klein1984huckel}, essential for obtaining the targeted chemical properties of new drugs. These findings suggest that InstGAN successfully produced new drug-like molecules with relatively high QED, logP, and SA scores.

Figures E.3 and E.4 in the appendix depict the curves of average chemical property values versus training steps for molecules generated in single- and multi-property optimization tasks. Chemical properties for both single- and multi-property optimization exhibited gradual increases over 5000 training steps. In single-property optimization, the independence of the three chemical properties resulted in distinct and noticeable score increases. Moreover, during the multi-property optimization process, the mutual constraints between properties led to similar changing trends in the values of the three properties.

\begin{figure*}[t]
\centering
\includegraphics[width=0.9\textwidth]{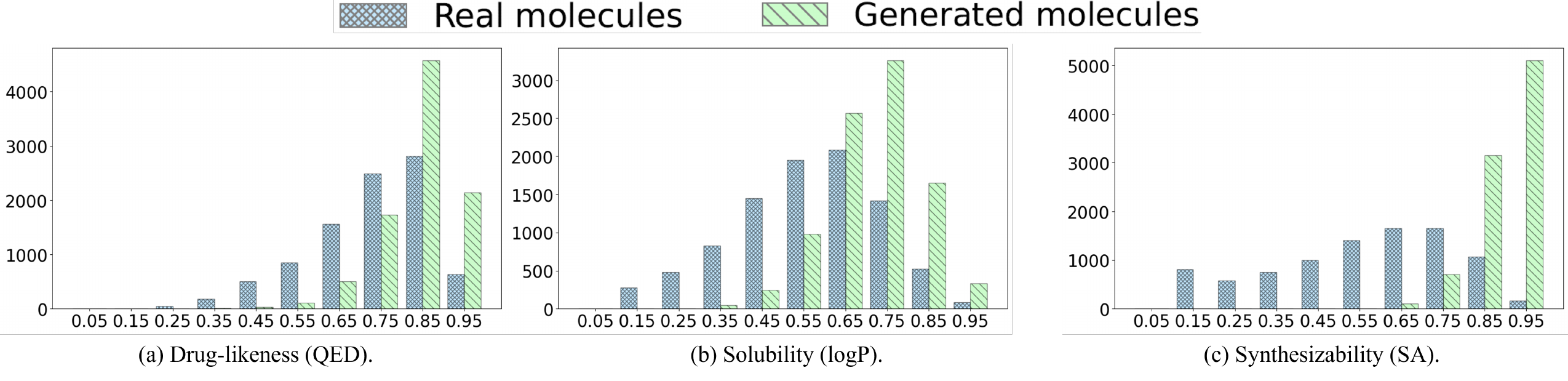}
\caption{Property distributions of generated molecules with multi-property optimization.}
\label{fig:multi_dis}
\end{figure*}

Fig. \ref{fig:multi_dis} and Figure E.5 in the Appendix show the property distributions of molecules generated with multi- and single-property optimization, respectively. Intuitively, in comparison to the molecule distributions in the training set (in blue), the chemical property distributions of the newly generated molecules (in green) shifted to the right overall. This suggests that InstGAN produced a greater number of new molecules with desirable properties. Furthermore, the property scores of molecules generated through multi-property optimization were marginally lower than those from single-property optimization.

This difference arises primarily because InstGAN had to concurrently consider the enhancement of three properties and several of which are inherently conflicting during multi-property optimization. Detailed results and trade-off analyses are provided in Appendix F.
In essence, multi-property optimization for molecular generation proves to be more challenging and intricate.The results demonstrated the effectiveness of InstGAN in chemical property optimization.

\subsection{Ablation Studies}
\label{subsec:ablation}
Table \ref{table:ablation} demonstrates the impact of various InstGAN variants on molecular generation performance. During the training process of InstGAN, we excluded GR, and MIE individually to create three distinct variants, namely, “w/o IR," “w/o GR," and “w/o MIE." These variants were then compared with InstGAN. In the “w/o IR" scenario, the calculation of instant reward was substituted with the MCTS-based RL algorithm. While MCTS enhanced GAN training stability with its extensive sampling, 
\begin{table}
\centering
\resizebox{0.4\textwidth}{!}{
\begin{tabular}{lccccc}\toprule
& Validity & Uniqueness & Novelty & Total  \\\hline
w/o IR & 97.80 & 70.71 & 98.02 & 67.79 \\
w/o GR & 95.96 & \colorbox[gray]{0.9}{98.72} & 99.66 & 94.41 \\
w/o MIE & \colorbox[gray]{0.9}{98.39} & 96.50 & 99.54 & 94.51 \\\hline
\textbf{InstGAN} & 97.56 & 98.47 & \colorbox[gray]{0.9}{99.73} & \colorbox[gray]{0.9}{95.81} \\\toprule
\end{tabular}}
\caption{Effect of different variants of InstGAN.}
\label{table:ablation}
\end{table}
resulting in relatively high validity (97.80\%), its computational complexity, stemming from the large number of samples, constrains its applicability in lengthy sequences. In the case of “w/o GR", the validity was the lowest (95.96\%, compared to 97.56\% of InstGAN). Given that the GR involves the global information of a SMILES string, it provides additional sequence-related information for generating subsequent tokens in the molecular auto-regression process, thereby contributing to the validity improvement. MIE, by smoothing the probabilities of generating tokens, enhances diversity in sampling tokens with non-maximum probabilities. Consequently, in the “w/o MIE" scenario, the generated molecular distribution exhibited the lowest uniqueness (96.50\%, compared to 98.47\% of InstGAN). InstGAN, incorporating IR, GR, and MIE, 
achieved the highest total score of 95.81\%. 

Additionally, Tables G.1, G.2, G.3, and G.4 display the effects of $\lambda$ on the performance. InstGAN can be applied to extensive chemical databases, optimizing chemical properties while retaining a low computational cost.

Integrating an actor-critic RL with instant and global rewards is crucial to the success of InstGAN. When compared to MCTS sampling, our strategy is more computationally eifficient, and its fine-grained feedback over SMILES sub-strings provides informative gradients at the early training stage mitgating the gradient vanishing problem that often hinders discrete GANs. The same mechanism also stabilizes the learning process: in the integration setup of GAN and RL, the generator simultaneously minimizes KL divergence and maximizes JS divergence, producing conflicting gradients \cite{saxena2021generative}. Injecting entropy through token-level rewards and MIE can balance the two objectives and aligns with RL findings that entropy regularization facilitates exploration and robust learning \cite{ahmed2019understanding}. As a result, InstGAN achieves more stable convergence, improved sample quality, and performance on par with SOTA models.
\subsection{Case Studies}
\label{subsec:case}
In the case studies, our goal is to generate molecules with high QED and a significant affinity for DRD2 within the ChEMBL database. This pursuit is critical for identifying potential drug candidates distinguished by enhanced drug-like properties and targeted therapeutic effects.  This significance is underscored by the wealth of available experimental bioactive data, providing a robust foundation for advancing promising compounds in drug discovery.

\begin{figure}
\centering
\includegraphics[width=0.9\hsize]{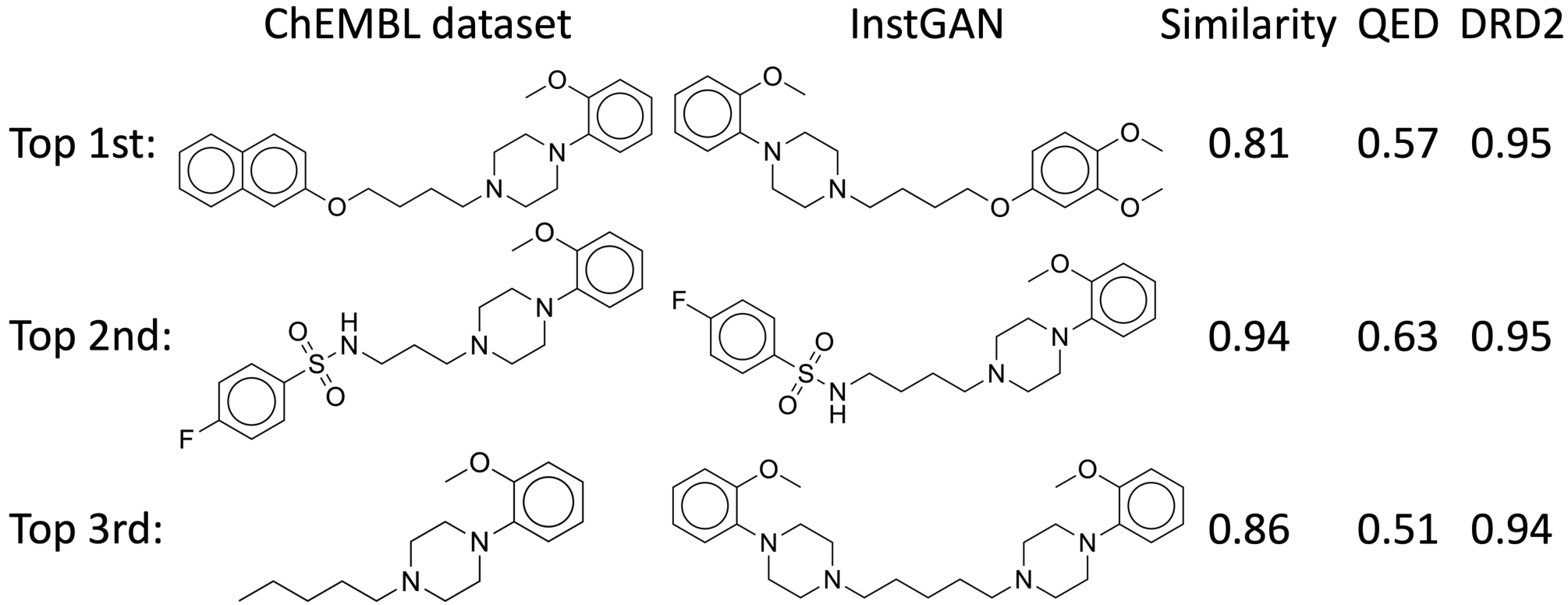}
\caption{Comparison of the generated molecules with high DRD2 scores and similar approved drugs.}
\label{fig:case_top3}
\end{figure}

Table H.1 in Appendix H assesses the performance of QED and DRD2 properties, demonstrating that the QED and DRD2 scores change with the corresponding weights. Furthermore, InstGAN enhanced the bioactivity of the generated molecules to 97.21\%. Additionally, we selected a QED and DRD2 weight ratio of (0.3, 0.7) and generated bioactive molecules. Figure \ref{fig:case_top3} compares the generated molecules with high DRD2 scores to similar approved drugs. These Top-3 molecules generated by InstGAN have high QED and DRD2 scores and are highly similar to approved drugs in the ChEMBL database, proving the effectiveness of InstGAN.

\section{Conclusion}
\label{sec:conclusion}
This study introduced InstGAN for generating molecules with multi-property optimization from scratch. Unlike MCTS-based RL algorithms, we employed an actor-critic RL algorithm for the efficient computation of IR and GR, resulting in reduced computation time and stabilized molecular generation quality. Additionally, the inclusion of MIE was used to alleviate the mode collapse problem and promote diversity in molecular generation. The experimental results demonstrated that InstGAN achieves comparable performance to SOTA baseline models and efficiently generates molecules with single- and multi-property optimization. 

InstGAN has two main limitations. First, the number of critics increases with the number of chemical properties that need to be optimized, which leads to an increase in the training cost. Second, the inclusion of additional hyperparameters, such as $\lambda$ and $W_{C_i}$, requires manual tuning, posing a challenge for fine-tuning. In future work, we will explore solutions to address these challenges.

\section*{Acknowledgements}
This research was partially supported by the International Research Fellow of Japan Society for the Promotion of Science (Postdoctoral Fellowships for Research in Japan [Standard]) and KAKENHI (20K11830, 25K15130), Japan.

\bibliographystyle{named}
\bibliography{ijcai25}

\end{document}